\documentclass[twocolumn,prl]{revtex4}  
\usepackage{graphicx}
\usepackage{epsfig}
\usepackage{amsmath}
\usepackage{amsfonts}
\usepackage{amssymb}
\usepackage{float}
\usepackage{psvectorian}

\begin{document}

\title{Isomorphism between Maximum Lyapunov Exponent and Shannon's Channel Capacity}

\author{Gerald Friedland$^{1,2}$, Alfredo Metere$^{3*}$}

\affiliation{$^1$ Department of Electrical Engineering and Computer Science, University of California - Berkeley, Soda Hall, Berkeley, CA - 94720, USA\\$^2$ Engineering Directorate,
  Lawrence Livermore National Laboratory, 7000 East Avenue L-795,
  Livermore, CA - 94550, USA\\$^3*$ Physical and Life Science Directorate,
  Lawrence Livermore National Laboratory, 7000 East Avenue L-367,
  Livermore, CA - 94550, USA\\metere1@llnl.gov}


\begin{abstract}
We demonstrate that the Maximum Lyapunov Exponent for computable dynamical systems is isomorphic to the maximum capacity of a noiseless, memoryless channel in a Shannon communication model. The isomorphism allows the understanding of Lyapunov exponents in the simplified terms of Information Theory, rather than the traditional definitions in Chaos Theory. This work provides a bridge between fundamental physics and Information Theory to the mutual benefit of both fields. The result suggests, among other implications, that machine learning and other information theory methods can be successfully employed at the core of physics simulations.
\end{abstract}

\date{\today}

\maketitle

\begin{figure*}
\center
\includegraphics[width=16cm]{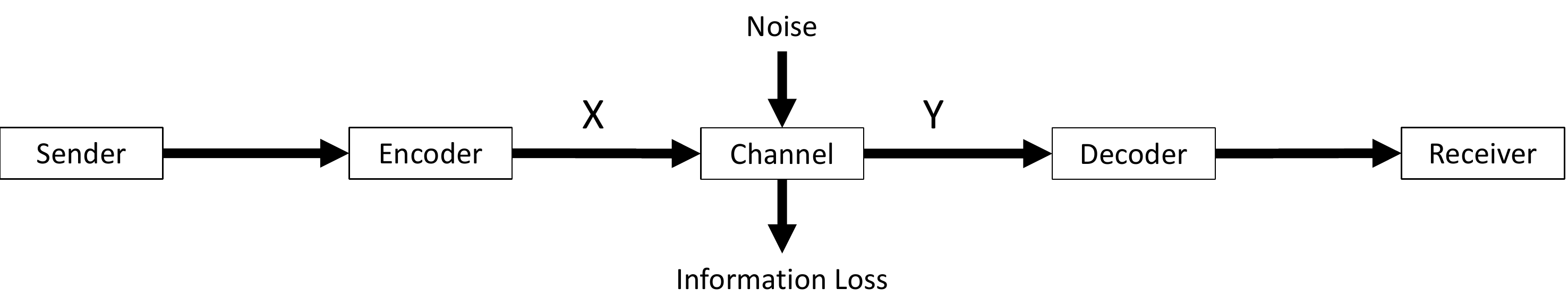}
\caption{Schematic of Shannon's communication model. $X$ and $Y$ represent the channel's encoded input and output respectively.}
\label{fig1}
\end{figure*}
Information Theory is a relatively new field of study introduced by Shannon in 1948, providing mathematical understanding of how information can be measured, stored and transmitted between senders and receivers, meant in the most generic way possible.
Today, Information Theory represents the foundation of telecommunications, signal processing and machine learning, e.g., Artificial Neural Networks are defined in Information Theory as universal encoders \cite{mackay2003}.\\
Shannon's formulation~\cite{shannon2001mathematical} defines information as the message, and its flow as a channel-mediated communication model.
As Fig.~\ref{fig1} depicts, information is the content of a message that is sent by a sender to a receiver, passing through a channel.
In order for the channel to pass the message, it needs to be encoded, and in order for the receiver to understand the message, it must be decoded.
All these elements, known as the Shannon Communication Model, can be meant either very generally or very specifically without a change in the paradigm.
For example, consider a speech system consisting of a person $A$, the \textit{sender}, who wants to communicate his/her idea, a spoken \textit{message}, to person $B$, the \textit{receiver}.
For the success of the communication, person $A$, using the vocal cords and the mouth-nose complex, must \textit{encode} the message into codified sound waves: the verbal language.
The air bulk between person $A$ and person $B$ represents the \textit{channel} through which the encoded message propagate as sound waves, and the ear of person $B$ is the \textit{decoder} able to transform the codified sound waves back into electrical signals that the brain can use to infer person $A$'s ideas contained in the message.
The communication happened successfully if person $B$'s idea is an accurate enough inference of person $A$'s idea. \\
The channel in this scenario is considered to be \textit{noisy}, which means that the encoded message gets distorted by the channel, and therefore the chances that the message cannot be correctly transmitted are not negligible.
In this example, the noise can either come from the channel itself dampening the amplitude of the signal. 
Noise can also be introduced in the channel by polluting it with additional undesired background noise consisting of other sound waves.
In contrast, an ideal, \textit{noiseless} channel will not be affected by any noise.
However, noiseless channels can only be ideal, and they are studied to understand the theoretical limits of how real channels function.
Channels could also have memory or be memoryless. 
A memory channel's behavior is influenced by the input, while memoryless channels exhibit input-independent behavior.\\

The most important measure to establish the quality of communication is the \textit{channel capacity} ~\cite{shannon2001mathematical}, which measures how much information can be successfully communicated per unit of time by a certain channel, given a certain message.
A channel is characterized by its channel capacity $C$:
\begin{align}
C & = \sup_{p_X(x)} I (X;Y), \label{eqn:chancapstat}
\end{align}
where $X$ and $Y$ are, respectively, the encoded input and output of the channel; $p_X(x)$, is the marginal distribution \cite{mackay2003}, which is the probability distribution of the variables contained in the subset $x$ for a given set $X$; and $I(X;Y)$ is the mutual information \cite{mackay2003}, which measures the mutual dependence between the $X$ and $Y$.
Information Theory models rely on probability theory, and they are generic.
Such a stochastic formulation stems from certain fundamental assumptions, e.g., that information can be treated statistically and that the Central Limit Theorem (CLT), which can be summarized by saying that when the size of a set of independent distributions tends to infinity, the convolution of all the distributions in the set converges to a Gaussian distribution.
In the stochastic definition $X$ and $Y$ are true random variables that are not necessarily correlated, hence the statistical mutual dependence formulation in Eq.~\ref{eqn:chancapstat}. 
However, in reality the message always needs time to transition from input to output through a channel, hence we can define the channel as a time-dependent function $N(t)$ that maps the input $X$ to the output $Y$ as follows~\cite{mackay2003,shannon2001mathematical}:
\begin{align}
\frac{Y}{X} = N(t). \label{eqn:xysignal}
\end{align}
Such relationship already suggests that a channel communication could in principle be treated as the time-evolution of a dynamical system in physics, where $X$ and $Y$ would represent respectively the initial and final state, and $N(t)$ the equation of motion for such equivalent dynamical system.\\
Thanks mainly to the work of Shannon and Hartley, several models have been reformulated in a more practical form to describe the behavior of real-world communication systems. 
For example, the channel capacity (Eq.~\ref{eqn:chancapstat}) can be reformulated using the Shannon-Hartley theorem \cite{shannon2001mathematical}, for a noiseless, memoryless channel as:
\begin{align}
C & = \lim_{t \to \infty} \frac{1}{t} \log_2 \left ( \frac{Y}{X} \right ),  \label{eqn:chancap}
\end{align}
where $t$ is the sampling period.
The $\log_2(x)$ is an encoding operator for $x$ and it returns the real-valued number of bits necessary to represent the message, and if $t$ is expressed in seconds, then the dimensions of $C$ are:
\begin{align}
C = [s^{-1}] [\text{bits}] \label{eqn:dimchancap}.
\end{align}
This formulation exposes that $C$ necessarily depends on the sampling frequency $t^{-1}$ and on the input of the channel $X$, while the ratio $Y/X$ is determined by the nature of the channel (see Eq.~\ref{eqn:xysignal}). \\
The amount of information passing through the channel is called \textit{signal} \cite{shannon2001mathematical}, and it can be defined as:
\begin{align}
S = \log_2(Y) - \log_2(X) = \log_2 \left ( \frac{Y}{X} \right ) , \label{eqn:signal}
\end{align}
which is the bit-wise difference between output and input. The sampling frequency $t^{-1}$ is commonly called bandwidth.\\

Among the main contributors to the foundation of modern Chaos Theory, which is a branch of mathematics stemming from the study of non-linear dynamical systems in Physics, Kolmogorov and Sinai thoroughly investigated the link between Information Theory and Physics.
Their work resulted into the stochastic interpretability of physical phenomena, especially at a quantum level, commonly referred to as \textit{Stochastic Physics} or \textit{Stochastic Mechanics} ~\cite{vulpiani2009}.
However, because of the generality and the abstraction of the used probabilistic formulations, their theoretical work needs non-trivial interpretation and adaptation to be used for real-world physical systems.\\
Another important contributor to Chaos Theory is Lyapunov, who developed several, more practical tools for characterizing the behavior of non-linear dynamical systems.
More specifically, the chaotic behavior of computable Hamiltonian dynamical systems can be characterized by the Lyapunov Exponents (LE), a set of quantities that estimate the rate of separation between infinitesimally close trajectories \cite{vulpiani2009}.
For example, let a 3D Hamiltonian dynamical system of arbitrary form consisting of $N$ particles, defining a $6N$-dimensional phase space. For each particle, we can calculate two trajectories, starting from two distinct points in phase space initially separated by a distance $|\delta \mathbf{Z}(t_0)|$ at time $t_0$. 
The separation distance $|\delta \mathbf{Z}(t)|$ at any time $t > t_0, ~ t \to \infty$ can be calculated as follows:
\begin{align}
\lim_{t \to \infty}{| \delta \mathbf{Z}(t) |} = |\delta \mathbf{Z}(t_0)| \lim_{t \to \infty} e^{\lambda t}  , \label{eqn:lyapunovpre}
\end{align}
where $\lambda$ is the Lyapunov exponent for each particle. We can reformulate Eq.~\ref{eqn:lyapunovpre} to highlight $\lambda$ as follows~\cite{vulpiani2009}:
\begin{align}
\lambda = \lim_{t \to \infty} \frac{1}{t} \ln \left( \frac{ |\delta \mathbf{Z}(t)| }{ |\delta \mathbf{Z}(t_0)| } \right). \label{eqn:lyapunov2}
\end{align}
Therefore, for our system there will be a set of $6N$ Lyapunov exponents, known as \textit{Lyapunov spectrum}. Because our system is symplectic, the volume of the phase space is preserved, thus resulting in a total of $3N$ negative and $3N$ positive Lyapunov exponents, the sum of them being zero. 
The presence of positive Lyapunov exponents is a necessary but not sufficient condition for a dynamical system to be defined as chaotic. However, in chaotic systems, the larger the LEs, the faster the chaotic system will become unpredictable.
For this reason, the most important Lyapunov exponent is the largest, commonly referred in literature as Maximum Lyapunov Exponent (MLE), and it is obtained by assuming that the initial separation is infinitesimally small, as follows~\cite{vulpiani2009}:
\begin{align}
\lambda_{M} & = \lim_{t \to \infty} ~\lim_{| \delta \mathbf{Z}(t_0) | \to 0} ~ \frac{1}{t} \ln \left ( \frac{ |\delta \mathbf{Z}(t)|}{| \delta \mathbf{Z}(t_0) |} \right ), \label{eqn:lyapunov}
\end{align}
where $\lambda_M$ is the MLE.
Lyapunov exponents can represent the exponential separation between trajectories of the same system for two infinitely close initial states, but could also represent the separation between the real trajectory of a dynamical system and its discrete, finite-state, computable surrogate \cite{vulpiani2009}.\\

The similarity between Eq.~\ref{eqn:chancap} and Eq.~\ref{eqn:lyapunov2} is already evident. In fact, the two formulations are isomorphic. 
In this article we want to demonstrate such isomorphism, list and briefly discuss what we believe to be the most important implications of this relationship both in Information Theory and in Physics.\\

\section{Isomorphism}
Let $\Psi$ be an noiseless, memoryless, channel conveying an initial separation distance $X = | \delta \mathbf{Z}(t_0) |$ at time $t_0$ to a final separation distance  $ Y = | \delta \mathbf{Z}(t) |$ at time $t$, with $t \to \infty$ in a Hamiltonian dynamical system:
\begin{align}
\Psi : X \to Y  \label{eqn:channelmap} .
\end{align}

The signal $S$ sent by the channel can be quantified, according to Eq. \ref{eqn:signal}, as follows:
\begin{align}
S & = \log_2 \left ( \frac{Y}{X} \right ) = \log_2 \left ( \frac{| \delta \mathbf{Z}(t) |}{| \delta \mathbf{Z}(t_0) |} \right )  \label{eqn:signal1} .
\end{align}
Hence, the channel capacity $\lambda_\Psi$ of the channel $\Psi$ can be calculated as:
\begin{align}
\lambda_\Psi & = \lim_{t \to \infty} \frac{1}{t} \log_2 \left ( \frac{Y}{X} \right ) \\
             & = \lim_{t \to \infty} \frac{1}{t} \log_2 \left ( \frac{| \delta \mathbf{Z}(t) |}{| \delta \mathbf{Z}(t_0) |} \right ) . \label{eqn:chancap2}
\end{align}
The forms of the Eq. \ref{eqn:lyapunov2} and \ref{eqn:chancap2} are similar. Since an isomorphism \cite{weisstein} exists between logarithms:
\begin{align}
\log_x(a) \cong \frac{\log_y(a)}{\log_y(x)}~~ \forall \{ a,~x \in \mathbb{R}^1~~|~~ a,~x > 0  \} \label{eqn:isolog} ,
\end{align}
by induction, we identify the following isomorphism:
\begin{align}
\lambda \cong \lambda_\Psi  \label{eqn:isomorph} .
\end{align}
$\square$

\section{Discussion}
To the best of our knowledge, the isomorphism presented above has never been formalized before.
Each implication is deep enough to deserve the writing of a separate manuscript. Therefore, we will only list and briefly comment on each of the most important consequences of this isomorphism, both from Chaos Theory and Information Theory perspectives.\\

We would like to refresh the reader about Artificial Neural Networks (ANNs) being defined as universal encoders in Information Theory \cite{mackay2003}. The reported isomorphism implies that ANNs can be thought of and can be used as non-linear, chaotic-approximant memory storage units, implementable in hardware, able to learn with enough accuracy any dynamical system trajectory. ANNs can therefore be used to predict states of the system that were not part of the input dataset, without explicitly calculating the system's equations of motion.\\
The Lyapunov Exponent (or channel capacity) of a dynamical system can be seen as the amount of bits necessary to keep track of the time evolution of the system's states with satisfactory accuracy.
Because of that, this isomorphism represents the first step towards the creation of a new field of investigation based on the adoption of Deep Learning and other Machine Learning methods to accurately simulate dynamical systems trajectories without explicitly solving the equations of motion.
Such approach represents a revolution in the field of computational physics.
It is well known \cite{vulpiani2009, vulpiani1, vulpiani2, caiani1997} that Information Theory models are comparable to 1D recursive maps in Chaos Theory. This isomorphism implies that the computation of any $N-$dimensional dynamical system can always be reduced to compute 1D recursive maps without loss of generality: we can call this new definition as  \textit{principle of computer-representability invariance} of a physical system.
From such principle it immediately follows that the complexity reduction of data representation translates into increased encoding complexity.
This suggests that lossless compression could be seen as a complexity-preserving transformation. \\

Another important implication of this isomorphism is that a new metric, the Maximum Channel Capacity $C_M$, can be defined for a noiseless, memoryless channel as follows:
\begin{align}
C_M & = \lim_{t \to \infty} ~ \lim_{X \to 0} \frac{1}{t} \log_2 \left ( \frac{Y}{X} \right )  \label{eqn:maxchancap} ,
\end{align}
which we proved hereby to be isomorphic to Eq.~\ref{eqn:lyapunov}.
Differently from what was already reported by Shannon (Eq. \ref{eqn:chancap}), we introduced the limit $\lim_{X \to 0}$, which has the physical meaning of an infinitesimally small message to be sent through a channel. 
In this form, we actually measure the maximum channel capacity needed for a particular channel independently on the size of the input, rather than a specific, input-dependent channel capacity, thus defining the upper limit of possible capacities for the range of accepted inputs.\\
Kolmogorov-Sinai (KS) Entropy, a very important and well-known metric for the study of dynamical systems, can be defined as the sum of the positive Lyapunov exponents of a dynamical system~\cite{vulpiani2009, vulpiani1, vulpiani2, misha1998}. This isomorphism reveals that analogously, KS Entropy can also be measured as the total bit rate corresponding to the sum of the positive channel capacities of a communication system.
Hence, this result extends the validity of Lyapunov analysis to all known systems treatable by Information Theory, although we should remark that Lyapunov exponents have been previously used, not surprisingly, to determine the channel capacity of finite-state Markov channels and memory channels \cite{holliday2006,pfister2003}.\\
We are also aware of the work done by M. Ebeid about the stochastic relationship between LEs and Information Theory limits, using Control Theory models~\cite{hani}.
Our work validates and significantly extends the results previously reported by M. Ebeid. We prove that the LE is not simply equal to the communication rate in some broad family of channels, as M. Ebeid reported, but the LE is in fact isomorphic to the communication rate. We additionally propose a fully deterministic formulation, in net contrast with the exclusively stochastic approach adopted by M. Ebeid, suggesting that systems studied by information theory can also be legitimately described deterministically and viewed in physics as dynamical systems.\\

\section{Conclusions}
In conclusion, this isomorphism constitutes a decisive proof that Information Theory models can be adopted for the analysis and characterization of dynamical systems simulation data, including the possibility of rationally using Machine Learning to accelerate the sampling of the phase space of Hamiltonian systems. \\

\section{Acknowledgements}
This work has been supported in part by the Joint Design of Advanced Computing Solutions for Cancer (JDACS4C) program established by the U.S. Department of Energy (DOE) and the National Cancer Institute (NCI) of the National Institutes of Health and was performed under the auspices of the U.S. Department of Energy by  Lawrence Livermore National Laboratory under Contract DE-AC52-07NA27344. IM release number LLNL-TR-733786. It was also partially supported by a Lawrence Livermore Laboratory Directed Research \& Development grants (17-ERD-096, 17-SI-004, and 18-ERD-021). Any findings and conclusions are those of the authors, and do not necessarily reflect the views of the funders. We want to warmly thank Prof. Angelo Vulpiani, Dr. Aaron Wilson, and Dr. Sachin S. Talathi for suggestions, and Dr. Jeffrey Hittinger for illuminating suggestions and funding support.


\end{document}